\documentclass[12pt]{article}
\usepackage{amsmath}
\usepackage{graphicx}
\usepackage{subfigure}
\usepackage{hyperref}

\begin{document}

\author{Ernst Trojan and George V. Vlasov \and \textit{Moscow Institute of Physics
and Technology} \and \textit{PO Box 3, Moscow, 125080, Russia}}
\title{Interaction and heat exchange in two-component relativistic fluid composite}
\maketitle

\begin{abstract}
A model of two-component relativistic fluid is considered, and the thermal
nature of coupling between the fluid constituents is outlined. This thermal
coupling is responsible for non-ideality of the fluid composite where the
components are not fully independent. The interaction between particles is
reflected only in the equation of state of each component, but it deals
nothing with the coupling between the fluid components and does not
influence the hydrodynamic motion. A general form of two-fluid decomposition
is formulated for arbitrary interacting system. 
\end{abstract}

\section{Introduction}

Two-component and multi-component relativistic fluid systems are often
considered in various problems of nuclear physics and astrophysics. As a
rule, the matter is described in the frames of ideal fluid mechanics.
However, the coupling between fluid components may give rise to non-ideality
of the fluid composite. Relativistic hydrodynamics of non-ideal
multi-component fluid \cite{Carter89} involves complicated equations but it
allows to operate with more precise theory. Particularly, the theory of
two-component relativistic fluid has found important applications to
hydrodynamics of relativistic superfluidity \cite{Israel81,KL82} and nuclear
hydrodynamics of proton-neutron systems \cite{CJ2003}. The stress-energy
tensor of a two-component fluid \cite{CL95a} 
\begin{equation}
T_{\nu \sigma }=n_\nu \mu _\sigma +s_\nu \Theta _\sigma +P\,g_{\nu \sigma }
\label{t}
\end{equation}
includes the particle number current $n_\nu $\ and the flux of chemical
potential $\mu _\sigma $\ corresponding to the first component, and the
particle number current $s_\nu $\ and the flux of chemical potentia $\Theta
_\sigma $\ corresponding to the second component. In a relativistic
superfluid the variables $n_\nu $ and $\mu _\sigma $\ belong to the ''cold''
component, while the entropy density current $s_\nu $\ and the temperature
flux $\Theta _\sigma $\ belong to the ''warm'' component.

The vectors in (\ref{t}) are not independent but obey relationship 
\begin{equation}
\left( 
\begin{array}{c}
n^\nu \\ 
s^\nu
\end{array}
\right) =\left( 
\begin{array}{cc}
F & Q \\ 
Q & G
\end{array}
\right) \left( 
\begin{array}{c}
\mu ^\nu \\ 
\Theta ^\nu
\end{array}
\right)  \label{mat}
\end{equation}
with the coefficients 
\begin{equation}
F=\frac 1\mu \frac{\partial P}{\partial \mu }\qquad G=\frac 1\Theta \frac{%
\partial P}{\partial \Theta }\qquad Q=\frac 1z\frac{\partial P}{\partial z}
\label{coe}
\end{equation}
because the pressure $P$, in general, depends on three variables 
\begin{equation}
\mu ^2=-\mu ^\nu \mu _\nu \qquad \Theta ^2=-\Theta ^\nu \Theta _\nu \qquad
z^2=-\mu ^\nu \Theta _\nu  \label{mtz}
\end{equation}
Substituting (\ref{coe}) in (\ref{t}) we write the stress-energy tensor in
the form 
\begin{equation}
T_{\nu \sigma }=F\mu _\nu \mu _\sigma +G\Theta _\nu \Theta _\sigma +2Q\mu
_\nu \Theta _\sigma +P\,g_{\nu \sigma }  \label{t1}
\end{equation}

The vanishing coefficient $Q=0$ will imply ideality of the two-fluid system
whose stress-energy tensor (\ref{t1}) will be written in the form 
\begin{equation}
T_{\nu \sigma }=F\mu _\nu \mu _\sigma +G\Theta _\nu \Theta _\sigma
+P\,g_{\nu \sigma }  \label{t00}
\end{equation}
or 
\begin{equation}
T_{\nu \sigma }=\mu nu_\nu u_\sigma +\Theta sv_\nu v_\sigma +P\,g_{\nu
\sigma }  \label{t000}
\end{equation}
It is a model of two ideal fluids without coupling. Velocities $u_\nu $ and $%
v_\nu $ can be different but there is no interference between them.

So, non-zero coefficient $Q$ implies that the motion of fluid components is
not fully independent. The pressure $P$ depends on the cross term $z$. This
fact reflects dependence on the relative velocity between the fluid
components 
\begin{equation}
w=\sqrt{1-\frac{\mu ^2\Theta ^2}{z^4}}  \label{w}
\end{equation}
In the light of (\ref{coe}), it results in non-zero coefficient $Q$ that
implies coupling between the components and non-ideality of the two-fluid
system: vectors $n_\nu $ and $\mu _\nu $ are not collinear (as well as $%
s_\nu $ and $\Theta _\nu $).

The coupling between the fluid components should not be mixed with the
inter-particle interaction. The first component can be an ideal gas of free
particles, the second component can be an ideal gas of free particles or
thermal excitations (quasi-particles), but the fluid components are coupled
as soon as the pressure $P$ depends on the relative velocity $w$ (\ref{w}).
It may look as an artifact of a plain mathematical trick without solid
physical background: as soon as the relative flow $w$ is introduced, the
thermodynamical parameters will contain functional dependence on $w$.
However, this idea was successfully developed in the Landau two-fluid model
of superfluid helium and it was confirmed in experiments \cite{LL87}.
Although the source of coupling between the ''cold'' and ''warm'' components
of relativistic superfluid is a consequence of interaction between the
particles \cite{KL82}, dependence on the relative velocity $w$ is not
evident in the field-theoretical approach to interacting many-particle
system \cite{Israel81,Israel89}, and it had been a subject of dispute for a
while \cite{KL82,Israel82}.

Researchers consider both variants -- either a mixture of two ideal fluids,
or a two-fluid composite where the coupling between components implies
non-ideality of the whole fluid system. The first model (without coupling)
is simple and it is often applied in astrophysics and nuclear physics. The
latter model (with coupling) is much more complicated and it promises to
give comprehensive description of hydrodynamic motion, although it is rarely
used on account of its complexity. Of course, it is very desirable to take
into account the role of coupling when we consider a strongly-interacting
medium, like superfluid or nuclear matter.

In the present paper we discuss the nature of coupling between the
components of a two-fluid composite and emphasize that it arises from the
heat exchange between them. The interaction between particles in a quantum
many-body system will also result in non-ideality of the continuous medium.
It is necessary to clarify what factor plays dominant role in the
hydrodynamics of a two-component fluid: the interaction between particles or
the thermal contact between fluid components?

Standard relativistic units $c_{light}=\hbar =k_B=1$\ are used in the paper.

\section{Thermal coupling between components}

Taking Equation (\ref{t1}), we immediately determine the energy density of a
two-component fluid \ 
\begin{equation}
E=T_0^0=F\mu ^0\mu _0+G\Theta ^0\Theta _0+2Q\mu _0\Theta _0+P\,g_0^0
\label{e}
\end{equation}
Non-ideality of the two-fluid composite, is due to non-zero term $Q$. In a
relativistic superfluid this is introduced at the macroscopic level of
hydrodynamics \cite{CL95a}. However, its appearance is a consequence of
internal processes in superfluid system \cite{Zhang2003} and it implies
dependence on the relative velocity between the ''cold'' and ''warm''
component $w$, that is confirmed in experimental research \cite{LL87}.\ The
similar term $Q$ will appear in the stress-energy tensor of an arbitrary two-component
fluid if its components are not fully independent \cite{Carter89}. What
process is responsible for non-ideality of this two-fluid composite? In order to
understand how this non-ideality arises, let us turn from classical
description to the quantum level.

For an arbitrary classical fluid system with the stress-energy tensor (\ref
{t1}) and the energy density $E$ (\ref{e}) we can recognize the Hamiltonian $%
\widehat{H}$ that determines the ground state $\left| \Phi \right\rangle $
according to the stationary Schr\"odinger equation 
\begin{equation}
 \widehat{H}\,\left| \Phi \right\rangle =\mathrm{E}_g\left| \Phi
\right\rangle  \label{dft}
\end{equation}
with the ground-state energy 
\begin{equation}
\mathrm{E}_g=\mathrm{\ }\left\langle \Phi \right| \widehat{H}\left| \Phi
\right\rangle =\int T_0^0\,d^3r  \label{dft2}
\end{equation}
Let us present the field variables $\mu _\nu $ and $\Theta _\nu $ in the
form \cite{V97b} \ 
\begin{equation}
\hat \mu _\nu \left( x\right) =\frac 1X\sum\limits_k\left\{ e_\nu \left(
k\right) \,\hat a_k\,e^{-ikx}+e_\nu ^{*}\left( k\right) \,\hat a_k^{\dagger
}\,e^{ikx}\right\}  \label{mu}
\end{equation}
\begin{equation}
\hat \Theta _\nu \left( x\right) =\frac 1Y\sum\limits_k\left\{ e_\nu \left(
k\right) \,\hat b_k\,e^{-ikx}+e_\nu ^{*}\left( k\right) \,\hat b_k^{\dagger
}\,e^{ikx}\right\}  \label{q}
\end{equation}
with a unit polarization vector $e_\nu $ and form-factors $X$ and $Y$, while
the annihilation operators $\hat a_k$ and $\hat b_k$ satisfy commutation
relations 
\begin{equation}
\left[ \hat a_k\,,\hat a_p^{\dagger }\right] =\delta _{kp}\qquad \qquad
\left[ \hat b_k\,,\hat b_p^{\dagger }\right] =\delta _{kp}  \label{bos1}
\end{equation}
and 
\begin{equation}
\left[ \hat a_k\,,\hat a_p\right] =\left[ \hat b_k\,,\hat b_p\right] =\left[
\hat a_k^{\dagger }\,,\hat a_p^{\dagger }\right] =\left[ \hat b_k^{\dagger
}\,,\hat b_p^{\dagger }\right] =0  \label{bos2}
\end{equation}

Substituting (\ref{mu}), (\ref{q}) in (\ref{e}), we find the Hamiltonian 
\begin{equation}
\widehat{H}=\int \widehat{T}_0^0\,d^3=V\sum\limits_k\left\{ \frac{2F_k}{X^2}%
\left( \hat a_k^{\dagger }\hat a_k+\frac 12\right) +\frac{2G_k}{Y^2}\left(
\hat b_k^{\dagger }\hat b_k+\frac 12\right) +\frac{2Q_k}{XY}\left( \hat
a_k^{\dagger }\hat b_k+\hat b_k^{\dagger }\hat a_k\right) \right\} +O
\label{h1}
\end{equation}
where $V$ is the volume of the system, $F_k$, $G_k$ and $Q_k$ are Fourier
images of $F\left( x\right) $, $G\left( x\right) $ and $Q\left( x\right) $,
while $O$ includes the terms with zero average in the light of (\ref{bos1})-(%
\ref{bos2}). Expression 
\begin{equation}
\hat N_k^a=\hat a_k^{\dagger }\hat a_k  \label{n1}
\end{equation}
is no more than the operator of number of particles with energy 
\begin{equation}
\varepsilon _k^a=\frac{2F_k}{X^2}  \label{e1}
\end{equation}
and momentum $k$. Expression 
\begin{equation}
\hat N_k^b=\hat b_k^{\dagger }\hat b_k  \label{n2}
\end{equation}
is the operator of number of particles with energy 
\begin{equation}
\varepsilon _k^b=\frac{2G_k}{Y^2}  \label{e2}
\end{equation}
and momentum $k$. Quantity 
\begin{equation}
\lambda _k=\frac{2Q_k}{XY}  \label{xi}
\end{equation}
also plays the role of energy. It reveals the fact of coupling between the
components because $\xi _k=0$ is equivalent to $Q=0$, and the Hamiltonian of
this ideal system 
\begin{equation}
\widehat{H}_0=\sum\limits_k\left\{ \varepsilon _k^a\left( \hat a_k^{\dagger
}\hat a_k+\frac 12\right) +\varepsilon _k^b\left( \hat b_k^{\dagger }\hat
b_k+\frac 12\right) \right\}  \label{hh00}
\end{equation}
corresponds to a two-component ideal fluid with the stress-energy tensor (%
\ref{t00}) or (\ref{t000}).

The Hamiltonian (\ref{hh00}) is already written in a diagonalized form. In
order to diagonalize the Hamiltonian of non-ideal fluid (\ref{h1}) let us
apply the following transformation 
\begin{equation}
\hat a_k=\cos \eta _k\,\hat \alpha _k+\sin \eta _k\,\hat \beta _k  \label{tr}
\end{equation}
\begin{equation}
\hat b_k=\cos \theta _k\,\hat \beta _k+\sin \theta _k\,\hat \alpha _k
\label{tr1}
\end{equation}
According to (\ref{bos1})-(\ref{bos2}), the operators $\hat \alpha _k$ and $%
\hat \beta _k$ also satisfy commutation relations 
\begin{equation}
\left[ \hat \alpha _k\,,\hat \alpha _p^{\dagger }\right] =\delta _{kp}\qquad
\qquad \left[ \hat \beta _k\,,\hat \beta _p^{\dagger }\right] =\delta _{kp}
\label{bos11}
\end{equation}
\begin{equation}
\left[ \hat \alpha _k\,,\hat \alpha _p\right] =\left[ \hat \beta _k\,,\hat
\beta _p\right] =\left[ \hat \alpha _k^{\dagger }\,,\hat \alpha _p^{\dagger
}\right] =\left[ \hat \beta _k^{\dagger }\,,\hat \beta _p^{\dagger }\right]
=0  \label{bos22}
\end{equation}
Then, the Hamiltonian (\ref{h1}) will be presented in a diagonalized form 
\begin{eqnarray}
\widehat{H} &=&W_0+\sum\limits_k\left\{ \left( \varepsilon _k^a\cos ^2\eta
_k+\varepsilon _k^a\sin ^2\theta _k+2\xi _k\cos \eta _k\sin \theta _k\right)
\hat \alpha _k^{\dagger }\hat \alpha _k+\right.  \nonumber \\
&&\ \ \left. \qquad +\left( \varepsilon _k^b\sin ^2\eta _k+\varepsilon
_k^b\cos ^2\theta _k+2\xi _k\sin \eta _k\cos \theta _k\right) \hat \beta
_k^{\dagger }\hat \beta _k\right\}  \label{hmd1}
\end{eqnarray}
where 
\begin{equation}
W_0=\frac 12\sum\limits_k\left\{ \varepsilon _k^a+\varepsilon _k^b\right\}
\label{e00}
\end{equation}
is the zero-point vibration energy, and parameters of transformation (\ref
{tr})-(\ref{tr1}) must satisfy condition 
\begin{equation}
2\lambda _k\left( \cos \eta _k\cos \theta _k+\sin \eta _k\sin \theta
_k\right) +\varepsilon _k^s\sin \left( 2\eta _k\right) +\varepsilon _k^n\sin
\left( 2\theta _k\right) =0  \label{diag}
\end{equation}
Particularly, at $\lambda _k=0$ condition (\ref{diag}) is simplified 
\begin{equation}
\varepsilon _k^a\sin (2\eta _k)+\varepsilon _k^b\sin (2\theta _k)=0
\label{diag0}
\end{equation}
and its solution is $\sin \eta _k=\sin \theta _k=0$ that implies no more
than 
\begin{equation}
\hat a_k=\hat \alpha _k\qquad \hat b_k=\hat \beta _k  \label{diag1}
\end{equation}
because the Hamiltonian (\ref{hh00}) is already given in a diagonalized form.

Thus, condition (\ref{diag}) imposed on transformation (\ref{tr})-(\ref{tr1}%
) determines a link between parameters $\eta _k$ and $\theta _k$. However,
it is not enough for their ultimate definition. An additional condition 
\begin{equation}
\varepsilon _k^0=\varepsilon _k^a\left( \cos ^2\eta _k+\sin ^2\theta
_k\right) =\varepsilon _k^b\left( \sin ^2\eta _k+\cos ^2\theta _k\right)
\label{diagx}
\end{equation}
determines exact $\eta _k$ and $\theta _k$ that allow to present the
Hamiltonian (\ref{hmd1}) in the following universal form 
\begin{equation}
\widehat{H}=\widehat{H}_{ideal}+\sum\limits_k\omega _k\left( \hat \alpha
_k^{\dagger }\hat \alpha _k-\hat \beta _k^{\dagger }\hat \beta _k\right)
\label{h2}
\end{equation}
where 
\begin{equation}
\widehat{H}_{ideal}=W_0+\sum\limits_k\left( \varepsilon _k^0+\Delta
_k\right) \left( \hat \alpha _k^{\dagger }\hat \alpha _k+\hat \beta
_k^{\dagger }\hat \beta _k\right)  \label{hh01}
\end{equation}
is the Hamiltonian of ideal system, while 
\begin{equation}
\omega _k=\lambda _k\sin \left( \eta _k-\theta _k\right)  \label{omxi}
\end{equation}
and 
\begin{equation}
\Delta _k=\lambda _k\sin \left( \eta _k+\theta _k\right)  \label{dxi}
\end{equation}
The energy shift $\Delta _k$ reflect no qualitative difference from the
ideal system with $Q=0$ and $\lambda _k=0$ whose Hamiltonian is 
\begin{equation}
\widehat{H}_0=W_0+\sum\limits_k\varepsilon _k^0\left( \hat \alpha
_k^{\dagger }\hat \alpha _k+\hat \beta _k^{\dagger }\hat \beta _k\right)
\label{hh02}
\end{equation}
So, there is no qualitative difference between two systems with
Hamiltonian (\ref{hh01}) and (\ref{hh02}). They both describe an ideal fluid. 
As soon as nonzero $Q\neq 0$ appears in the stress-energy tensor (\ref{t1}%
), it implies non-ideality, and the relevant Hamiltonian (\ref{h2}) includes
additional nontrivial terms with $\hat \alpha _k^{\dagger }\hat \alpha
_k-\hat \beta _k^{\dagger }\hat \beta _k$. To understand their role let us
appeal to the formalism of thermo field dynamics \cite{UMT82,MNU85}. Let us
introduce the operation of tilde-conjugation 
\begin{equation}
\widehat{\tilde \alpha }_k=\widehat{\beta }_k  \label{tilda}
\end{equation}
For operator 
\begin{equation}
\widehat{h}=\sum\limits_k\omega _k\hat \alpha _k^{\dagger }\hat \alpha _k
\label{dh1}
\end{equation}
its tilde-conjugated counterpart will be 
\begin{equation}
\widehat{\tilde h}=\sum\limits_k\omega _k\widehat{\tilde \alpha }_k^{\dagger
}\widehat{\tilde \alpha }_k =\sum\limits_k\omega _k\hat \beta _k^{\dagger
}\hat \beta _k  \label{dh2}
\end{equation}
and we can rewrite the Hamiltonian (\ref{h2}) in the following form 
\begin{equation}
\widehat{H}=\widehat{H}_{ideal}+\widehat{h}-\widehat{\tilde h}  \label{h3}
\end{equation}
while the Hamiltonian $\widehat{H}_{ideal}$ is invariant under
tilde-conjugation 
\begin{equation}
\widehat{H}_{ideal}=W_0+\sum\limits_k\left( \varepsilon _k^0+\Delta
_k\right) \widetilde{\left( \hat \alpha _k^{\dagger }\hat \alpha _k+\widehat{%
\tilde \alpha }_k^{\dagger }\widehat{\tilde \alpha }_k \right)} =
W_0+\sum\limits_k\left( \varepsilon _k^0+\Delta _k\right) \left( \widehat{%
\tilde \alpha }_k^{\dagger }\widehat{\tilde \alpha }_k+\hat \alpha
_k^{\dagger }\hat \alpha _k\right)  \label{h4}
\end{equation}
If we introduce thermal doublets 
\begin{equation}
\widehat{A}_k=\left( 
\begin{array}{c}
\hat \alpha _k \\ 
\widehat{\tilde \alpha }_k^{\dagger }
\end{array}
\right) \qquad \,\widehat{\bar A}_k=\left( 
\begin{array}{cc}
\hat \alpha _k^{\dagger } & -\widehat{\tilde \alpha }_k^{\dagger }
\end{array}
\right)  \label{dou}
\end{equation}
which satisfy commutation relations 
\begin{equation}
\left[ \widehat{A}_k,\widehat{\bar A}_p\right] =\delta _{kp}  \label{d2}
\end{equation}
then, the Hamiltonian (\ref{h3}) will be presented so 
\begin{equation}
\widehat{H}=\widehat{H}_{ideal}+\sum\limits_k\omega _k\left( \,\widehat{\bar
A}_k\,\widehat{A}_k+1\right)  \label{h5}
\end{equation}
or 
\begin{equation}
\widehat{H}=\widehat{H}_{ideal}^{\prime }+\sum\limits_k\omega _k\left( \,%
\widehat{\bar A}_k\,\widehat{A}_k+\frac 12\right)  \label{h6}
\end{equation}
where 
\begin{equation}
\widehat{H}_{ideal}^{\prime }=W_0^{\prime }+\sum\limits_k\left( \varepsilon
_k^0+\Delta _k\right) \left( \hat \alpha _k^{\dagger }\hat \alpha _k+\hat
\beta _k^{\dagger }\hat \beta _k\right)  \label{hz}
\end{equation}
and 
\begin{equation}
W_0^{\prime }=W_0+\frac 12\sum\limits_k\omega _k=\frac
12\sum\limits_k\left\{ \varepsilon _k^a+\varepsilon _k^b+\omega _k\right\}
\label{zero}
\end{equation}
Again, the shift of the zero-point vibration energy (\ref{zero}) reflects no
qualitative difference between the Hamiltonians (\ref{hz}) and (\ref{hh02}).
However, the last term in the Hamiltonian (\ref{h6}) makes essential
difference from the ideal system with the Hamiltonian (\ref{hh02}).

We can take the Hamiltonian of non-ideal fluid in various forms, applying
the tilde-conjugation (\ref{tilda}) and thermal doublets (\ref{dou}).
However, no transformation can reduce this Hamiltonian to the form which
corresponds to the Hamiltonian of an ideal fluid. The last term in (\ref{h6}%
) is responsible for the coupling between the components. We can consider
the Hamiltonian of non-ideal fluid in the form (\ref{h3}), where the last
two terms are responsible for the coupling. In the Hamiltonian taken in the
form (\ref{h2}) the last terms, again, include the coupling. Now we can
explain the very nature of the coupling between the components of non-ideal
fluid. It is resulted from exchange of thermal excitations in the process of
absorption of a quantum with the energy $\omega _k$ and momentum $k$ (action
of operator $\hat \alpha _k$), or in the process of annihilation of a hole
with the energy $-\omega _k$ and the same momentum $k$ (action of operator $%
\hat \beta _k$). The excitation and the hole with respect to the zero energy
level of the ideal fluid $E_0^{\prime }$ (\ref{zero}). If the components are
isolated and there is no heat exchange between them, then, $\omega _k\equiv
0 $ and it is necessary to be $\widehat{h}=\widehat{\tilde h}=0$\ that
corresponds to an ideal fluid. Thus, the heat exchange is the source of
coupling between the fluid components. The similar interpretation is known
in hydrodynamics of superfluid helium where collective motion of thermal
excitations is responsible for hydrodynamic behavior of the whole system 
\cite{LL87}. Although this link between thermal excitations and macroscopic
properties of the continuous medium is intuitively evident, we have proved
it in the explicit analysis with formulas (\ref{h2}), (\ref{h3}) and (\ref
{h6}) for the first time.

However, it is still unclear what is the role of interaction between
particles.

\section{Interaction between particles}

The pressure $P$ and energy density $E$ of a many-particle system are
defined by standard formulas \cite{Kapusta89} 
\begin{equation}
PV=-\Theta \ln Z  \label{p}
\end{equation}
and 
\begin{equation}
EV=\Theta ^2\, \frac{\partial \ln Z}{\partial \Theta }  \label{ep}
\end{equation}
where the statistical sum $Z$ is determined by formula 
\begin{equation}
\ln Z\left( \varepsilon _p,\mu ,\Theta \right) =\mp \frac{\gamma V}{\left(
2\pi \right) ^3}\int d^3p\ln \left[ 1\pm \exp \left( \frac{\mu -\varepsilon
_p}\Theta \right) \right]  \label{s0}
\end{equation}
with the upper and low sign corresponding to fermions and bosons, and $%
\varepsilon _p$ is the single-particle energy. For a system in motion at a
velocity $\vec w$ the statistical sum is determined by formula \cite{CL95a} 
\begin{equation}
\ln Z\left( \varepsilon _p^{\prime },\mu ,\Theta \right) =\mp \frac{\gamma V%
}{\left( 2\pi \right) ^3}\int d^3p\ln \left[ 1\pm \exp \left( \frac{\mu
-\varepsilon _p^{\prime }}\Theta \right) \right]  \label{s}
\end{equation}
and the distribution function is \cite{ST83} 
\begin{equation}
f_p\left( \Theta \right) =\frac 1{\exp \left( \varepsilon _p^{\prime }-\mu
\right) /\Theta \pm 1}  \label{ocu1}
\end{equation}
where 
\begin{equation}
\varepsilon _p^{\prime }=\varepsilon _p-\vec p\cdot \vec w  \label{sp}
\end{equation}
is the single-particle energy in the moving reference frame.

Now consider a two-component fluid composite. When there is no relative flow
between the components, the statistical sums of each component will be
determined by the same formula (\ref{s0}) in the co-moving reference frame,
or by the same formula (\ref{s}) in the laboratory reference frame. What
happens if relative motion between the fluid components occurs? Let us
consider this two-fluid system in the reference frame co-moving the first
component. Then, the statistical sum of the first component is defined by
formula (\ref{s0}), that can be written briefly so 
\begin{equation}
Z_a=Z\left( \varepsilon _p^a,\mu _a,\Theta \right)  \label{sa}
\end{equation}
The statistical sum of the second component is calculated, according to (\ref
{s}), in the reference frame co-moving the first component: 
\begin{equation}
Z_b=Z\left( \varepsilon _p^b-\vec p\cdot \vec w,\,\mu _b,\Theta \right)
\label{sb}
\end{equation}
where $\vec w$\ implies the relative velocity between the first and second
components. 

The total statistical sum of two-component system is 
\begin{equation}
Z=Z_{a\,}Z_b  \label{zab}
\end{equation}
Hence 
\begin{equation}
\ln Z=\ln \left( Z_{a\,}Z_b\right) =\ln Z_a+\ln Z_b  \label{zab2}
\end{equation}
and, according to (\ref{p}) and (\ref{ep}), the pressure and energy density
are additive quantities 
\begin{equation}
P=P_a+P_b  \label{pp}
\end{equation}
\begin{equation}
E=E_a+E_b  \label{ee}
\end{equation}

How to complete a two-fluid decomposition for a system of interacting
particles? It is convenient to apply the ideology of density functional
theory that considers a system of free particles under the action of
external self-consistent field as a model of real interacting system. This
ideology is developed in the nuclear mean-field approximation \cite{SW86}
where fluctuations of meson fields are neglected, and particles move
independent in the mean fields, which themselves are generated
self-consistently by the particles. This mean-field approximation allows to
split the statistical sum 
\begin{equation}
Z=\tilde ZZ^{\psi \,}  \label{sz}
\end{equation}
into a product of the statistical sum of mean fields $\tilde Z$ and
statistical sum of particles $Z^{\psi \,}$where the latter is defined
according to (\ref{s0}). A set of self-consistent mean-field equations
determine the effective mass $M_{*}$ and effective chemical potential $\mu
_{*}$ which depend on the constant of interaction. 
As a result of (\ref{p}), (\ref{ep}) and (\ref{sz}), the pressure and energy
density of the whole interacting system is 
\begin{equation}
P\left( \varepsilon _p^{*},\mu _{*},\Theta \right) =P_\psi \left(
\varepsilon _p^{*},\mu _{*},\Theta \right) +\tilde P\left( \varepsilon
_p^{*},\mu _{*},\Theta \right)   \label{p4}
\end{equation}
\begin{equation}
E\left( \varepsilon _p^{*},\mu _{*},\Theta \right) =E_\psi \left(
\varepsilon _p^{*},\mu _{*},\Theta \right) +\tilde E\left( \varepsilon
_p^{*},\mu _{*},\Theta \right)   \label{e4}
\end{equation}
where $P_\psi $ and $E_\psi $ imply contributions of the particles, and $%
\tilde P$ and $\tilde E$ are contributions of the mean fields, while 
\begin{equation}
\varepsilon _p^{*}=\sqrt{p^2+M_{*}^2}  \label{ef3}
\end{equation}

Again, following (\ref{p}), (\ref{ep}) and (\ref{sa})-(\ref{zab}), we
determine the pressure and energy of the first component as 
\begin{equation}
P_a=P\left( \varepsilon _p^{*a},\mu _{*}^a,\Theta \right) =P_\psi \left(
\varepsilon _p^{*},\mu _{*},\Theta \right) +\tilde P\left( \varepsilon
_p^{*},\mu _{*},\Theta \right)   \label{p5}
\end{equation}
\begin{equation}
E_a=E\left( \varepsilon _p^{*a},\mu _{*}^a,\Theta \right) =E_\psi \left(
\varepsilon _p^{*},\mu _{*},\Theta \right) +\tilde E\left( \varepsilon
_p^{*},\mu _{*},\Theta \right)   \label{e5}
\end{equation}
while the pressure and energy of the second component will be 
\begin{equation}
P_b=P\left( \varepsilon _p^{*b}-\vec p\cdot \vec w,\mu _{*}^b,\Theta \right)
=P_\psi \left( \varepsilon _p^{*b}-\vec p\cdot \vec w,\mu _{*}^b,\Theta
\right) +\tilde P\left( \varepsilon _p^{*b}-\vec p\cdot \vec w,\mu
_{*}^b,\Theta \right)   \label{p6}
\end{equation}
\begin{equation}
E_b=E\left( \varepsilon _p^{*b}-\vec p\cdot \vec w,\mu _{*}^b,\Theta \right)
=E_\psi \left( \varepsilon _p^{*b}-\vec p\cdot \vec w,\mu _{*}^b,\Theta
\right) +\tilde E\left( \varepsilon _p^{*b}-\vec p\cdot \vec w,\mu
_{*}^b,\Theta \right)   \label{e6}
\end{equation}
As soon as we know the thermodynamical functions of interacting system (\ref
{p4})-(\ref{e4}), we can define the thermodynamical functions of the two
components (\ref{p5})-(\ref{e5}) and (\ref{p6})-(\ref{e6}), thus making a
two-fluid decomposition (\ref{pp})-(\ref{ee}).

\section{Conclusion}

Non-ideality of the two-component superfluid at the macroscopic level is
expressed in additional coefficient $Q$ that appears in the stress-energy
tensor (\ref{t1}) and reflects its dependence on the relative flow $w$,
resulting to coupling between the fluid components. The analysis in the
frames of thermo field dynamics has revealed that the Hamiltonian contains
irreducible terms responsible for coupling and implying the heat exchange
between the components. This effect of thermal contact is taken into account
when the relative velocity between the components is introduced in the
thermodynamical functions of a two-fluid system.

One can consider the thermodynamical functions independent on the relative
velocity $w$, that corresponds to the vanishing coefficient $Q=0$. This
model describes a two-fluid system where the components are in full thermal
isolation. Of course, each component can be a strongly-interacting matter
rather than ideal gas.

Indeed, a two-fluid model with the heat exchange between the components
pertains better to superfluid systems because the ''cold'' and ''warm'' are
not independent, and interference between them is evident. On the other
hand, a two- or multi-component nuclear matter can be also described as an
ideal two-component fluid if no macroscopic relative flow is expected. It
can be a proton-neutron or quark-gluon system in equilibrium where only
infinitesimal relative motion between the components is admitted in the form
of spin-isospin sound.

The the essence of two-fluid model is expressed in the decomposition of
statistical sum (\ref{sa})-(\ref{zab}) that results in additive formulas for
the pressure (\ref{pp}) and energy density (\ref{ee}). The statistical
sum of the first component is calculated by standard formula (\ref{s0}),
while the statistical sum of the second component is determined by formula (%
\ref{s}) in the reference frame co-moving the first component. As a result,
the relative velocity between the components $w$ appears in the
thermodynamical functions.

When the pressure and energy density of interacting system is given (\ref{p4}%
)-(\ref{e4}), the two-fluid decomposition (\ref{pp})-(\ref{ee}) is completed
immediately according to formulas (\ref{p5})-(\ref{e5}) and (\ref{p6})-(\ref
{e6}).

Both effects of coupling between fluid constituents and interaction between
particles and taken into account. The interaction between particles is
responsible for non-ideality of the equation of state. It yields the
potential energy in addition to the kinetic energy of non-interacting
system. In the frames of mean-field approach it sets the effective mass $%
M_{*}$ and effective chemical potential $\mu _{*}$ that appear in formulas (%
\ref{p4})-(\ref{e6}). However, the hydrodynamic non-ideality of this fluid
system is not directly associated with the interaction between particles.

Without loss of generality, the two-fluid decomposition (\ref{pp})-(\ref{ee}%
) and (\ref{p5})-(\ref{e6}) can be also applied to superfluid systems, where
the potential energy of interaction includes additional anomalous fields
responsible for the condensation pressure and energy of the superfluid
ground state \cite{BL84}. For weakly interacting medium, like superfluid
helium, the second (or ''warm'') component is often considered as an ideal
gas of thermal excitations. However, such concept cannot be applied to
superfluid nuclear matter because the contribution of meson fields is not
small, and exact two-fluid decomposition (\ref{p5})-(\ref{e6}) should be
developed. It is the subject for further research.

The authors are grateful to Erwin Schmidt for discussions.

\end{document}